\documentclass[10pt, conference, letterpaper]{ IEEEtran}
\usepackage{cite}
\usepackage{amsmath,amssymb,amsfonts,mathtools,bm,amsthm}
\usepackage{graphicx}
\usepackage{textcomp}
\usepackage{xcolor}
\usepackage{bm}
\usepackage{algorithm}
\usepackage{algpseudocode}
\usepackage{booktabs}
\usepackage{array}
\usepackage{balance}
\usepackage[caption=false,font=footnotesize]{subfig}
\newcommand{\CN}{\mathcal{CN}}

\newcommand{\T}{\mathrm{T}}
\newcommand{\Hc}{\mathrm{H}}
\newcommand{\diag}{\operatorname{diag}}
\newcommand{\vecop}{\operatorname{vec}}

\begin{document}

\title{Joint Channel Estimation and Data Detection for Multi-LEO-Satellite Cell-Free OTFS Uplinks}

\author{\IEEEauthorblockN{Gangle~Sun\textsuperscript{1}, Tianhao~Liu\textsuperscript{2}, Jun~Tian\textsuperscript{3}, Xin~Cheng\textsuperscript{1}, Jian~Wu\textsuperscript{1}, Jinfang~Jiang\textsuperscript{1}, \\
Wenjin~Wang\textsuperscript{3},  Shi Jin\textsuperscript{3}, and Guangjie~Han\textsuperscript{1}}
	\textit{\textsuperscript{1}College of Information Science and Engineering, Hohai University, Changzhou, China}\\
    \textit{\textsuperscript{2}School of Integrated Circuits, Shanghai Jiao Tong University, Shanghai, China}\\
    \textit{\textsuperscript{3}National Mobile Communications Research Laboratory, Southeast University, Nanjing, China}\\
	\textit{email: sungangle@hhu.edu.cn, liutianhao26@sjtu.edu.cn, tianjun@seu.edu.cn, 20241011@hhu.edu.cn,}\\
	\textit{\{20262606, jiangjinfang\}@hhu.edu.cn, \{wangwj, jinshi\}@seu.edu.cn, and hanguangjie@gmail.com}\\
    }

\maketitle

\begin{abstract}
Cell-free networks formed by multiple low Earth orbit (LEO) satellites offer a promising architecture for ubiquitous connectivity, but their cooperative reception is challenged by link-dependent residual delays and Doppler shifts. This paper investigates joint channel estimation and data detection (JCEDD) for multi-LEO-satellite cell-free orthogonal time frequency space (OTFS) uplinks. The JCEDD problem is formulated as a structured bilinear inference problem involving link-specific sparse beam--delay--Doppler channels and a multiuser data vector. We develop a low-complexity hierarchical JCEDD receiver in which all satellites first perform local JCEDD, and their observations and local estimates are then aggregated at a central satellite for cooperative refinement. Computational complexity is reduced by restricting channel estimation to coarse-information-aided local beam--delay--Doppler regions and evaluating the required forward and adjoint operations in a matrix-free manner. Simulation results validate the channel-estimation accuracy and data-detection reliability of the proposed JCEDD receiver.
\end{abstract}


\section{Introduction}
Low Earth orbit (LEO) satellite constellations are expected to play an important role in sixth-generation (6G) wireless communication systems owing to their wide-area coverage, relatively short propagation distances, and ability to provide ubiquitous connectivity \cite{Wu2025distributed, Wang2025Toward, Yao2025Joint}. Nevertheless, reliable LEO satellite uplink reception remains challenging because of limited link budgets and the severe delay and Doppler effects induced by high satellite mobility. A cell-free architecture allows multiple spatially distributed LEO satellites to cooperatively receive the signals transmitted by the same ground users, thereby exploiting macro-diversity and complementary spatial observations. Meanwhile, orthogonal time frequency space (OTFS) modulation provides a structured delay-Doppler (DD)-domain representation of doubly selective channels. These features make multi-LEO-satellite cell-free OTFS a promising architecture for 6G ubiquitous connectivity.

Although coarse timing and frequency compensation is typically performed in multi-LEO-satellite systems, different satellite--user links generally retain distinct residual delays and Doppler shifts. Therefore, the receiver must estimate the link-specific doubly selective channels while detecting the common transmitted data from multiple satellite observations. The resulting joint channel estimation and data detection (JCEDD) problem is high-dimensional and bilinear, particularly when multiple satellites, antennas, users, and beam--delay--Doppler candidates are jointly considered. This motivates a cooperative JCEDD receiver that exploits multi-satellite observations while maintaining manageable computational and memory requirements.

\subsection{Related Work}

We review two closely related directions: multi-LEO-satellite cell-free transmission and OTFS-based receiver design for LEO and cell-free systems. Their limitations motivate the proposed framework.

\subsubsection{Multi-LEO-Satellite Cell-Free Transmission}
The cell-free concept was originally developed for terrestrial networks, in which spatially distributed access points jointly serve users without conventional cell boundaries \cite{Ngo2017Cell, Ngo2015Cell, Sun2026asynchronous}. Its extension to LEO satellite networks is attractive because multiple spatially separated satellites can receive the uplink signals of the same users and exploit macro-diversity. Existing studies have investigated distributed satellite multiple-input multiple-output (MIMO) architectures \cite{Abdelsadek2022Distributed}, centralized and federated coordination of satellite swarms \cite{Guidotti2024Federated}, dynamic satellite clustering \cite{Kim2025Cell}, and multi-LEO-satellite transmission with different waveforms and terminal configurations \cite{DAndrea2025Cell,Ramezani2026CellFreeMIMO}. These studies demonstrate the architectural potential of cell-free cooperation in LEO satellite networks.

Recent studies have further considered physical-layer reception in multi-LEO-satellite cell-free uplinks \cite{Shang2026Multi}. However, most closely related receiver designs focus on LEO-enabled massive random access. For example, \cite{Ying2023Quasi} develops a quasi-synchronous random-access receiver for massive-MIMO-based LEO satellite constellations, where edge satellites perform activity detection and channel estimation using a training-sequence-padded multicarrier structure, and a central node improves activity and data detection through multi-satellite cooperation. In \cite{Li2026angular}, the line-of-sight (LoS)-induced angular correlation among terrestrial--satellite links is exploited to improve grant-free active user detection and channel estimation in cooperative multi-satellite networks. In \cite{Xu2025MIMOBased}, the low-rank and row-sparse structures of the channel impulse-response matrix are used to develop a two-stage receiver for cooperative grant-free random access in MIMO-based multi-LEO-satellite Internet of Things (IoT) systems. These studies provide useful foundations for cooperative satellite reception, but they primarily exploit sporadic user activity and activity-domain sparsity rather than addressing JCEDD for scheduled multiuser uplinks.

\subsubsection{OTFS-Based Receiver Design for LEO and Cell-Free Systems}
OTFS represents information symbols in the DD domain, where doubly selective channels exhibit a more structured representation than in the conventional time-frequency domain \cite{Hadani2017OrthogonalIMS,Hadani2017OrthogonalWCNC,Wei2021Orthogonal}. Previous studies have investigated low-complexity OTFS implementation \cite{Farhang2018Low}, iterative data detection \cite{Raviteja2018Interference}, embedded-pilot channel estimation \cite{Raviteja2019Embedded}, diversity analysis \cite{Surabhi2019On}, massive-MIMO OTFS channel estimation \cite{Ramachandran2018MIMO,Liu2020Uplink,Shi2021Deterministic}, and sensing-aided receiver design \cite{Jiang2023Sensing}. These studies establish a useful receiver-design framework for high-mobility and doubly selective channels.

OTFS is particularly suitable for LEO satellite communications because its DD-domain representation can effectively characterize severe Doppler shifts and sparse high-mobility channels. In \cite{Shi2024OTFS}, the Doppler-resilient characteristics of OTFS for LEO links were discussed, together with a multi-satellite case study using DD-domain detection.  A systematic comparison between OTFS and orthogonal frequency-division multiplexing (OFDM) for multiuser LEO satellite communications was conducted in \cite{Liu2025OTFS}, where DD-domain resource allocation was also investigated. Beyond satellite communications, \cite{Ge2023OTFS} developed centralized and decentralized detectors for OTFS with sparse code multiple access (SCMA) in coordinated multi-point reception, illustrating the potential of combining distributed observations in high-mobility systems. OTFS signals emitted by LEO satellites have also been exploited for parameter estimation of range-migrating targets \cite{Ding2025Parameter}. Most existing communication-oriented receiver designs for LEO OTFS systems, however, focus on random-access scenarios.  Grant-free non-orthogonal multiple access (NOMA)-OTFS for LEO satellite Internet-of-Things was investigated in \cite{Zhou2023Active} for joint active user identification, channel estimation, and signal detection. Massive MIMO-OTFS random access was studied in \cite{Shen2022Random} and subsequently extended to joint device identification, channel estimation, and data detection in \cite{Shen2023Joint}.

OTFS has also been extended to cell-free systems. Terrestrial cell-free OTFS transmission was investigated in \cite{Mohammadi2022Cell}, while OTFS-based massive random access in cell-free massive-MIMO systems was considered in \cite{Hu2026A}. More closely related to this work, \cite{Shen2025Massive} studied cooperative multi-LEO-satellite OTFS random access and developed centralized and distributed receivers by exploiting angular-domain block sparsity and basis-expansion channel modeling. Although these studies demonstrate the benefits of OTFS and cooperative processing, JCEDD involving satellite-specific beam--delay--Doppler channels and data symbols jointly observed by multiple satellites remains insufficiently investigated.

\subsection{Motivation and Contributions}
Existing multi-LEO-satellite cell-free OTFS receiver designs are mainly developed for grant-free random access, where sporadic user activity is exploited for active-device identification and channel estimation. These formulations are not directly applicable to scheduled uplinks, where the active-user set is known and the main objective is to jointly recover the satellite-specific doubly selective channels and the transmitted data from multiple satellite observations. Moreover, fully centralized JCEDD may impose substantial computational, memory, and processing-delay requirements on the central satellite. Distributing part of the computation among the cooperating satellites before central refinement can alleviate this burden. Nevertheless, the joint consideration of multiple satellites, antennas, users, and beam--delay--Doppler candidates still leads to high-dimensional operations. These considerations motivate a low-complexity hierarchical local-to-central JCEDD receiver for multi-LEO-satellite cell-free OTFS uplinks.

The main contributions of this paper are listed as follows.
\begin{itemize}
\item We establish the signal model for scheduled multiuser multi-LEO-satellite cell-free OTFS uplinks under local single-satellite observations. Based on this model, JCEDD is formulated as a structured bilinear inference problem involving satellite--user-specific sparse beam--delay--Doppler channels and a stacked multiuser data vector jointly observed by all satellites.

\item We develop a hierarchical JCEDD receiver. Each satellite first performs local JCEDD using its own observation to obtain local channel and data estimates. The observations and local estimates are then aggregated at the central satellite, which jointly refines the satellite-specific channels and the common multiuser data using all available satellite observations.

\item We develop a low-complexity implementation by restricting channel estimation to coarse-information-aided local beam--delay--Doppler candidate regions. The required forward and adjoint operations are evaluated in a matrix-free manner, thereby reducing the channel search dimension and avoiding the explicit construction and storage of large sensing and equivalent-channel matrices.
\end{itemize}

\begin{figure*}[t]
    \centering
    \includegraphics[width=0.92\textwidth]{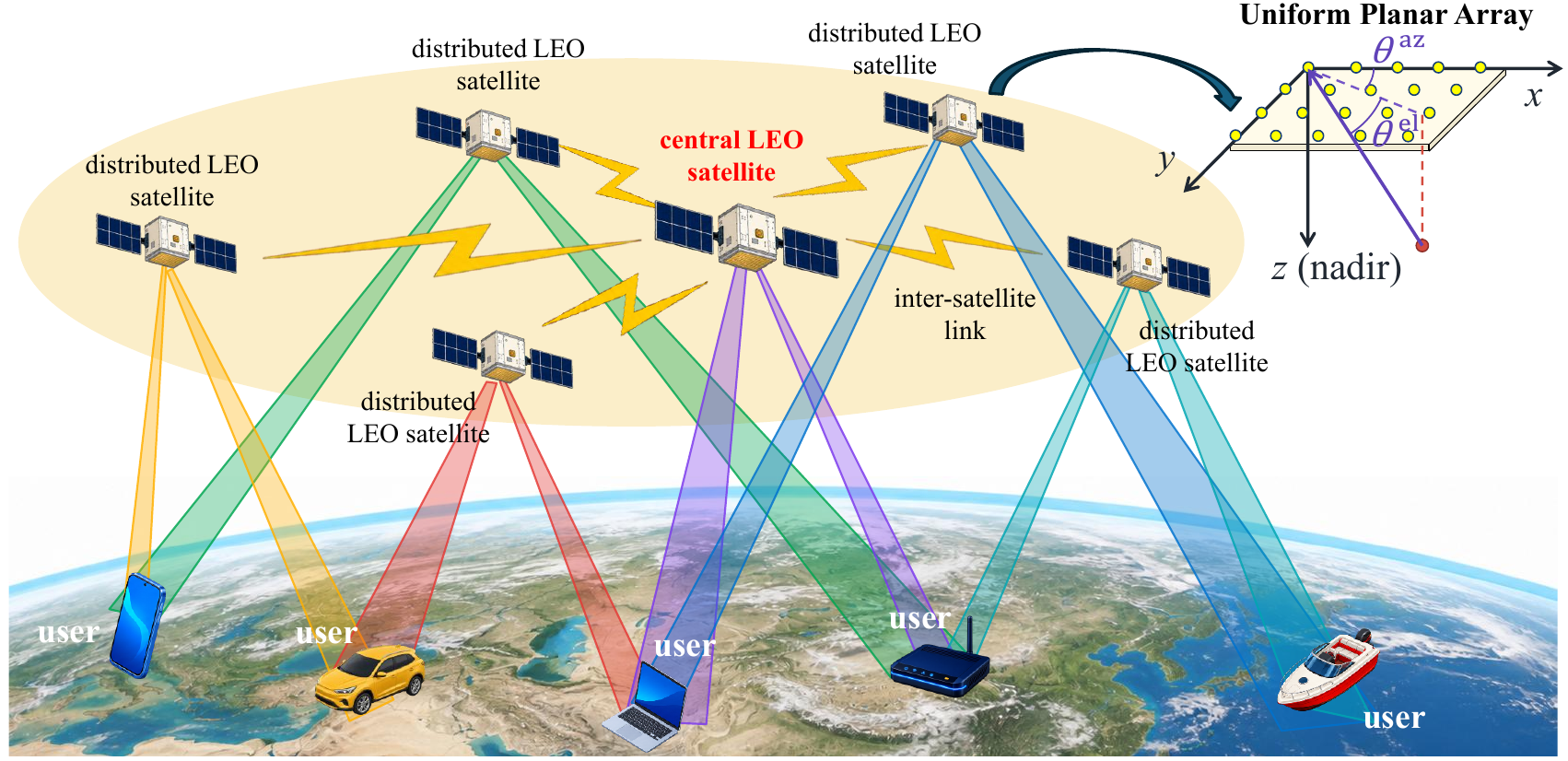}
    \caption{System architecture of the considered multi-LEO-satellite
    cell-free OTFS uplink, where multiple distributed LEO satellites
    cooperatively receive signals from ground users and aggregate their
    observations and local estimates at a central satellite for JCEDD.}
    \label{fig:system_scenario}
\end{figure*}

\subsection{Notation}
Matrices, column vectors, and sets are denoted by uppercase boldface letters, lowercase boldface letters, and calligraphic letters, respectively. The transpose and Hermitian transpose are denoted by $(\cdot)^\T$ and $(\cdot)^\Hc$, respectively. The Frobenius and Euclidean norms are denoted by $\|\cdot\|_F$ and $\|\cdot\|_2$, respectively.  The symbol $\mathbb{I}\{\cdot\}$ denotes the indicator function. The real and imaginary parts of a complex number are denoted by $\Re\{\cdot\}$ and $\Im\{\cdot\}$, respectively. The $N\times N$ identity matrix is denoted by $\mathbf{I}_N$, and the $a$th canonical basis vector is denoted by $\mathbf{e}_a$, with its dimension clear from the context. The operators $\otimes$, $\diag(\cdot)$, and $\vecop(\cdot)$ denote the Kronecker product, diagonalization, and vectorization, respectively. The complex Gaussian distribution with mean $\mathbf{m}$ and covariance matrix $\mathbf{C}$ is denoted by $\CN(\mathbf{m},\mathbf{C})$. The $L\times L$ unitary discrete Fourier transform (DFT) matrix is denoted by $\mathbf{F}_L$, whose $(r,s)$th entry is $[\mathbf{F}_L]_{r,s}=\frac{1}{\sqrt{L}}e^{-j2\pi rs/L}$ for $r,s=0,\ldots,L-1$.

\section{System Model}
\label{sec:system_model}
We consider the uplink of a multi-LEO-satellite cell-free system comprising $P$ receiving LEO satellites and $K$ scheduled single-antenna ground users, as illustrated in Fig.~\ref{fig:system_scenario}. Each satellite employs a uniform planar array (UPA) with $N_r=N_xN_y$ receive antennas. All satellites perform local processing, while one designated satellite additionally serves as the central processing satellite and aggregates the observations and local estimates from the other satellites.  The inter-satellite links are assumed to be error-free and sufficiently provisioned to convey the required observations and local estimates. Finite-capacity and quantized inter-satellite links are beyond the scope of this work.

Coarse timing and frequency compensation is performed using a reference satellite or nominal geometric information. Since the propagation conditions differ across satellite--user links, link-dependent residual delays and Doppler shifts remain after compensation. We therefore adopt OTFS modulation to represent the resulting doubly selective channels in the DD domain.

\subsection{OTFS Uplink Signal}
\label{subsec:otfs_model}
We consider reduced cyclic-prefix (CP) OTFS with rectangular pulses. One CP is appended to each OTFS frame and is assumed to cover the maximum residual path delay after coarse timing compensation. The DD grid contains $M$ delay bins and $N$ Doppler bins, with $Q=MN$. Let $\mathbf{X}_k\in\mathbb{C}^{M\times N}$ denote the DD-domain transmit matrix of user $k$. Its time-frequency (TF)-domain representation is
\begin{equation}
    \mathbf{X}_k^{\rm TF} = \mathbf{F}_M \mathbf{X}_k \mathbf{F}_N^{\mathsf H}.
    \label{eq:isfft}
\end{equation}

Using column-wise vectorization, define $\mathbf{x}_k=\vecop(\mathbf{X}_k)\in\mathbb{C}^{Q}$. The corresponding discrete-time OTFS waveform, excluding the CP, is
\begin{equation}
    \mathbf{s}_k = \mathbf{T}_{\rm OTFS}\mathbf{x}_k,
    \label{eq:otfs_modulation}
\end{equation}
where $ \mathbf{T}_{\rm OTFS} \triangleq \mathbf{F}_N^{\mathsf H}\otimes\mathbf{I}_M$.

The DD-domain transmit vector is decomposed as
\begin{equation}
    \mathbf{x}_k = \mathbf{p}_k + \mathbf{E}_k\mathbf{d}_k,
    \label{eq:pilot_data_decomposition}
\end{equation}
where $\mathbf{p}_k\in\mathbb{C}^{Q}$ contains the known pilot symbols and is zero at the data positions, $\mathbf{d}_k\in\mathcal{C}^{D_{\rm d}}$ contains the unknown data symbols, and $\mathbf{E}_k\in\{0,1\}^{Q\times D_{\rm d}}$ inserts $\mathbf{d}_k$ into the corresponding DD-domain positions. The pilot and data positions do not overlap, such that $\mathbf{E}_k^{\mathsf H}\mathbf{p}_k=\mathbf{0}$. All users transmit the same number $D_{\rm d}$ of data symbols. Unit-energy quadrature phase-shift keying (QPSK) is employed, with
\begin{equation}
    \mathcal{C}
    =
    \left\{
    \frac{\pm1\pm j}{\sqrt{2}}
    \right\}.
\end{equation}

The data vectors of all users are stacked as
\begin{equation}
    \mathbf{d}
    =
    \left[
    \mathbf{d}_1^{\mathsf T},
    \ldots,
    \mathbf{d}_K^{\mathsf T}
    \right]^{\mathsf T}
    \in\mathcal{C}^{D},
    \label{eq:stacked_data_vector}
\end{equation}
where $D=KD_{\rm d}$.

\subsection{Multi-Antenna DD-Domain Channel}
\label{subsec:physical_channel}

The channel from user $k$ to satellite $p$ is assumed to contain $L_{p,k}$ propagation paths. For path $\ell$, let $\alpha_{p,k,\ell}$, $\widetilde{\tau}_{p,k,\ell}$, and $\widetilde{\nu}_{p,k,\ell}$ denote the complex path gain, residual delay, and residual Doppler shift, respectively. These parameters are assumed constant within one OTFS frame.

Let
\begin{equation}
    \boldsymbol{\vartheta}_{p,k,\ell}
    =
    \left[
    \theta_{p,k,\ell}^{\rm az},
    \theta_{p,k,\ell}^{\rm el}
    \right]^{\mathsf T}
\end{equation}
denote the azimuth and elevation angles of arrival in the local coordinate system of satellite $p$, as illustrated in Fig.~\ref{fig:system_scenario}. The UPA is assumed to lie in the $x$--$y$ plane, and the elevation angle is measured from the array plane. The corresponding direction cosines are
\begin{align}
    u_x(\boldsymbol{\vartheta})
    &=
    \cos\theta^{\rm el}\cos\theta^{\rm az},\\
    u_y(\boldsymbol{\vartheta})
    &=
    \cos\theta^{\rm el}\sin\theta^{\rm az}.
\end{align}

Assuming half-wavelength antenna spacing, the steering vector along dimension $i\in\{x,y\}$ is
\begin{equation}
    \mathbf{a}_i
    \left(
    u_i(\boldsymbol{\vartheta})
    \right)
    =
    \frac{1}{\sqrt{N_i}}
    \left[
    1,
    e^{j\pi u_i(\boldsymbol{\vartheta})},
    \ldots,
    e^{j\pi(N_i-1)u_i(\boldsymbol{\vartheta})}
    \right]^{\mathsf T}.
    \label{eq:one_dimensional_steering_vector}
\end{equation}
With the $x$-dimension antenna index varying fastest, the UPA steering vector is
\begin{equation}
    \mathbf{a}_p(\boldsymbol{\vartheta})
    =
    \mathbf{a}_y
    \left(
    u_y(\boldsymbol{\vartheta})
    \right)
    \otimes
    \mathbf{a}_x
    \left(
    u_x(\boldsymbol{\vartheta})
    \right)
    \in\mathbb{C}^{N_r}.
    \label{eq:upa_steering_vector}
\end{equation}

Let $T_s$ denote the sampling interval. The signed frequency associated with the $m$th $Q$-point DFT bin is
\begin{equation}
    f_m
    =
    \begin{cases}
        \dfrac{m}{QT_s},
        & 0\leq m\leq
        \left\lfloor\dfrac{Q-1}{2}\right\rfloor,\\[2mm]
        \dfrac{m-Q}{QT_s},
        &
        \left\lfloor\dfrac{Q-1}{2}\right\rfloor<m\leq Q-1.
    \end{cases}
    \label{eq:signed_frequency}
\end{equation}

The circular-delay and Doppler operators are defined as
\begin{align}
    \mathbf{D}(\widetilde{\tau})
    &=
    \mathbf{F}_Q^{\mathsf H}
    \diag\!\left(
    e^{-j2\pi f_m\widetilde{\tau}}
    \right)_{m=0}^{Q-1}
    \mathbf{F}_Q,
    \label{eq:fractional_delay_operator}\\
    \mathbf{\Delta}(\widetilde{\nu})
    &=
    \diag\!\left(
    e^{j2\pi\widetilde{\nu}nT_s}
    \right)_{n=0}^{Q-1}.
    \label{eq:doppler_operator}
\end{align}
The circular-delay representation follows from the assumption that the CP covers all residual path delays. The corresponding DD-domain path operator is
\begin{equation}
    \mathbf{\Pi}
    (\widetilde{\tau},\widetilde{\nu})
    =
    \mathbf{T}_{\rm OTFS}^{\mathsf H}
    \mathbf{\Delta}(\widetilde{\nu})
    \mathbf{D}(\widetilde{\tau})
    \mathbf{T}_{\rm OTFS}
    \in\mathbb{C}^{Q\times Q}.
    \label{eq:dd_shift_spread_operator}
\end{equation}
This operator accommodates arbitrary residual delays and Doppler shifts. When $\widetilde{\tau}/T_s$ is an integer, $\mathbf{D}(\widetilde{\tau})$ reduces to a circular sample-shift operator~\cite{Sun2022massive}.

The equivalent antenna-DD-domain channel from user $k$ to satellite $p$ is
\begin{equation}
\begin{aligned}
    \mathbf{H}_{p,k}
    =
    \sum_{\ell=1}^{L_{p,k}}
    \alpha_{p,k,\ell}
    \Big[
    \mathbf{a}_p(\boldsymbol{\vartheta}_{p,k,\ell})
    \otimes
    \mathbf{\Pi}
    \left(
    \widetilde{\tau}_{p,k,\ell},
    \widetilde{\nu}_{p,k,\ell}
    \right)
    \Big],
\end{aligned}
\label{eq:antenna_dd_channel}
\end{equation}
where $\mathbf{H}_{p,k}\in\mathbb{C}^{N_rQ\times Q}$. The received DD-domain vector at satellite $p$ is
\begin{equation}
    \mathbf{y}_p
    =
    \sum_{k=1}^{K}
    \mathbf{H}_{p,k}\mathbf{x}_k
    +
    \mathbf{w}_p,
    \label{eq:antenna_domain_received_signal}
\end{equation}
where $\mathbf{w}_p \sim \mathcal{CN} \left(\mathbf{0},\sigma_p^2\mathbf{I}_{N_rQ} \right)$ is the noise vector at satellite $p$ with per-entry variance $\sigma_p^2$.

\subsection{Reduced Beam--Delay--Doppler Model}
\label{subsec:reduced_channel}

To exploit the angular sparsity of the multi-antenna LEO channels, define the two-dimensional DFT beam basis as
\begin{equation}
    \mathbf{U}_{\rm a}
    =
    \mathbf{F}_{N_y}^{\mathsf H}
    \otimes
    \mathbf{F}_{N_x}^{\mathsf H}
    \in\mathbb{C}^{N_r\times N_r}.
    \label{eq:beam_basis}
\end{equation}
The corresponding beam-domain observation at satellite $p$ is
\begin{equation}
    \bar{\mathbf{y}}_p
    =
    \left(
    \mathbf{U}_{\rm a}^{\mathsf H}
    \otimes
    \mathbf{I}_Q
    \right)
    \mathbf{y}_p.
    \label{eq:beam_domain_received_signal}
\end{equation}
A path whose spatial frequency lies on the DFT grid is concentrated in one beam bin, whereas an off-grid path generally spreads over several neighboring beam bins. Since $\mathbf{U}_{\rm a}$ is unitary, the transformed noise remains spatially white.

Based on coarse ephemeris, geometric, and synchronization information, the channel of link $(p,k)$ is restricted to the local candidate sets
\begin{align}
    \Omega_{p,k}^{\rm b}
    &=
    \left\{
    b_{p,k,a}
    \right\}_{a=1}^{A_{p,k}},\\
    \Omega_{p,k}^{\tau}
    &=
    \left\{
    \tau_{p,k,t}
    \right\}_{t=1}^{T_{p,k}},\\
    \Omega_{p,k}^{\nu}
    &=
    \left\{
    \nu_{p,k,v}
    \right\}_{v=1}^{V_{p,k}}.
    \label{eq:local_candidate_sets}
\end{align}
Here, $b_{p,k,a}$ is the $a$th candidate beam-bin index, while $\tau_{p,k,t}$ and $\nu_{p,k,v}$ are the candidate residual delays and Doppler shifts, respectively. These sets determine only the local search region; the associated channel coefficients remain unknown.

To retain all candidate beams required by satellite $p$, define
\begin{equation}
    \Omega_p^{\rm b}
    =
    \bigcup_{k=1}^{K}
    \Omega_{p,k}^{\rm b},
    \label{eq:union_candidate_beams}
\end{equation}
and $ B_p  = \left| \Omega_p^{\rm b} \right|$. Let $\mathbf{S}_p\in\{0,1\}^{B_p\times N_r}$ be the selection matrix whose rows extract the beam bins in $\Omega_p^{\rm b}$. The reduced beam-domain observation is
\begin{equation}
    \mathbf{y}_p^{\rm b}
    =
    \left(
    \mathbf{S}_p\otimes\mathbf{I}_Q
    \right)
    \bar{\mathbf{y}}_p
    \in\mathbb{C}^{B_pQ}.
    \label{eq:reduced_beam_observation}
\end{equation}
Since $\mathbf{S}_p\mathbf{S}_p^{\mathsf H}=\mathbf{I}_{B_p}$, the corresponding noise satisfies $\mathbf{w}_p^{\rm b}\sim\mathcal{CN}\left(\mathbf{0}, \sigma_p^2\mathbf{I}_{B_pQ} \right)$.

For each candidate tuple $(a,t,v)$, let $h_{p,k,a,t,v}\in\mathbb{C}$ denote the corresponding unknown beam--delay--Doppler coefficient. The associated atom is
\begin{equation}
    \mathbf{A}_{p,k,a,t,v}
    \triangleq
    \left(
    \mathbf{S}_p\mathbf{e}_{b_{p,k,a}}
    \right)
    \otimes
    \mathbf{\Pi}
    \left(
    \tau_{p,k,t},
    \nu_{p,k,v}
    \right)
    \in\mathbb{C}^{B_pQ\times Q},
    \label{eq:beam_delay_doppler_atom}
\end{equation}
where $\mathbf{e}_{b_{p,k,a}}\in\mathbb{C}^{N_r}$ is the corresponding canonical basis vector.

Collecting the coefficients over all beam, delay, and Doppler candidates gives $\mathbf{h}_{p,k}\in\mathbb{C}^{J_{p,k}}$, where $ J_{p,k} = A_{p,k}T_{p,k}V_{p,k}$ is the total number of candidate beam--delay--Doppler atoms for link $(p,k)$ and the Doppler index varies fastest in $\mathbf{h}_{p,k}$, followed by the delay and beam indices. The reduced equivalent channel is represented by the linear mapping
\begin{equation}
\begin{aligned}
    \mathbf{H}_{p,k}^{\rm r}
    \left[
    \mathbf{h}_{p,k}
    \right]
    \triangleq
    \sum_{a=1}^{A_{p,k}}
    \sum_{t=1}^{T_{p,k}}
    \sum_{v=1}^{V_{p,k}}
    h_{p,k,a,t,v}
    \mathbf{A}_{p,k,a,t,v}
    \in
    \mathbb{C}^{B_pQ\times Q}.
\end{aligned}
\label{eq:reduced_equivalent_channel}
\end{equation}

This candidate-dictionary model provides a sparse discretized representation of the physical channel. Because each physical link contains only a small number of propagation paths, only a small number of entries of $\mathbf{h}_{p,k}$ are expected to be significant. Off-grid paths may be represented by multiple neighboring beam--delay--Doppler atoms.

Under this representation, the reduced received-signal model is given by
\begin{equation}
    \mathbf{y}_p^{\rm b}
    =
    \sum_{k=1}^{K}
    \mathbf{H}_{p,k}^{\rm r}
    \left[
    \mathbf{h}_{p,k}
    \right]
    \mathbf{x}_k
    +
    \mathbf{w}_p^{\rm b}.
    \label{eq:reduced_received_model}
\end{equation}

Finally, the channel vectors of all users observed at satellite $p$ are stacked as
\begin{equation}
    \mathbf{h}_p
    =
    \left[
    \mathbf{h}_{p,1}^{\mathsf T},
    \ldots,
    \mathbf{h}_{p,K}^{\mathsf T}
    \right]^{\mathsf T}
    \in\mathbb{C}^{J_p},
    \label{eq:stacked_local_channel}
\end{equation}
where $J_p = \sum_{k=1}^{K} J_{p,k}$ denotes the total number of candidate beam--delay--Doppler atoms associated with satellite $p$.

\section{Low-Complexity Hierarchical JCEDD Receiver}
\label{sec:receiver}

The proposed hierarchical JCEDD receiver consists of parallel local JCEDD and central cooperative refinement. Each satellite first estimates its local beam--delay--Doppler channels and a local copy of the transmitted data using its own observation. The observations and local estimates are subsequently aggregated at the central satellite, which jointly refines the satellite-specific channels and the common multiuser data. Both stages operate on the reduced beam--delay--Doppler model and evaluate the required forward and adjoint operations without explicitly constructing large matrices.

\subsection{Local JCEDD}
\label{subsec:local_receiver}

Satellite $p$ jointly estimates its channel vector $\mathbf{h}_p$ and a local copy of the multiuser data vector $\mathbf{d}$ from $\mathbf{y}_p^{\rm b}$. For any $\mathbf{x}\in\mathbb{C}^{Q}$, define $\mathbf{\Phi}_{p,k}[\mathbf{x}]\in\mathbb{C}^{B_pQ\times J_{p,k}}$ as the sensing matrix whose columns are $\mathbf{A}_{p,k,a,t,v}\mathbf{x}$ over all candidate tuples $(a,t,v)$, ordered consistently with the entries of $\mathbf{h}_{p,k}$. It follows that
\begin{equation}
    \mathbf{\Phi}_{p,k}[\mathbf{x}]
    \mathbf{h}_{p,k}
    =
    \mathbf{H}_{p,k}^{\rm r}
    [\mathbf{h}_{p,k}]
    \mathbf{x}.
    \label{eq:equivalent_representations}
\end{equation}
These matrix forms are introduced only for notational convenience. In the implementation, the corresponding forward and adjoint operations are evaluated using fast Fourier transforms (FFTs), phase rotations, beam selection, and summations over the candidate atoms, without explicitly constructing $\mathbf{\Phi}_{p,k}[\mathbf{x}]$ or $\mathbf{H}_{p,k}^{\rm r}[\mathbf{h}_{p,k}]$.

Let $\mathcal{B}=\operatorname{conv}(\mathcal{C})$ denote the convex hull of the QPSK constellation, i.e.,
\begin{equation}
    \mathcal{B}
    =
    \left\{
    z\in\mathbb{C}:
    |\Re\{z\}|\leq\frac{1}{\sqrt{2}},
    \;
    |\Im\{z\}|\leq\frac{1}{\sqrt{2}}
    \right\}.
    \label{eq:qpsk_box}
\end{equation}
The local JCEDD problem at satellite $p$ is formulated as~\cite{Sun2025deep}
\begin{equation}
\begin{aligned}
    \min_{\mathbf{h}_p,\,\mathbf{d}\in\mathcal{B}^{D}}
    \,
    &
    \frac{
    \|\mathbf{r}_p[\mathbf{h}_p,\mathbf{d}]\|_2^2
    }{2\sigma_p^2}
    +
    \mu_h
    \sum_{k=1}^{K}
    \|\mathbf{h}_{p,k}\|_1
    -
    \frac{\lambda_d}{2}
    \|\mathbf{d}\|_2^2,
\end{aligned}
\label{eq:local_joint_problem}
\end{equation}
where the local residual term is defined as
\begin{equation}
    \mathbf{r}_p[\mathbf{h}_p,\mathbf{d}]
    \triangleq
    \mathbf{y}_p^{\rm b}
    -
    \sum_{k=1}^{K}
    \mathbf{H}_{p,k}^{\rm r}[\mathbf{h}_{p,k}]
    \left(
    \mathbf{p}_k+\mathbf{E}_k\mathbf{d}_k
    \right).
    \label{eq:local_residual_function}
\end{equation}
The $\ell_1$ regularizer promotes sparse channel coefficients, while the box constraint and negative quadratic regularizer encourage the relaxed data estimates to approach the QPSK constellation.

Problem \eqref{eq:local_joint_problem} is nonconvex because of the bilinear channel--data coupling and the negative quadratic regularizer. We therefore apply a forward-backward splitting (FBS)-type proximal-gradient method to obtain a stationary-point estimate. The iterations are initialized as
\begin{equation}
    \mathbf{h}_p^{(0)}=\mathbf{0},
    \qquad
    \mathbf{d}^{(0)}=\mathbf{0}.
    \label{eq:local_initialization}
\end{equation}
At iteration $i$, define
\begin{align}
    \mathbf{x}_k^{(i)}
    &\triangleq
    \mathbf{p}_k+\mathbf{E}_k\mathbf{d}_k^{(i)},
    \label{eq:iterative_transmitted_signal}\\
    \mathbf{r}_p^{(i)}
    &\triangleq
    \mathbf{r}_p
    [\mathbf{h}_p^{(i)},\mathbf{d}^{(i)}].
    \label{eq:iterative_residual}
\end{align}

Using the convention $\nabla_{\mathbf{z}}f=2\partial f/\partial\mathbf{z}^{*}$, the channel blocks are updated as
\begin{equation}
\begin{aligned}
    \mathbf{h}_{p,k}^{(i+1)}
    =
    \mathcal{S}_{\gamma_{h,p}^{(i)}\mu_h}
    \left(
    \mathbf{h}_{p,k}^{(i)}
    +
    \frac{\gamma_{h,p}^{(i)}}{\sigma_p^2}
    \mathbf{\Phi}_{p,k}
    [\mathbf{x}_k^{(i)}]^{\mathsf H}
    \mathbf{r}_p^{(i)}
    \right),
\end{aligned}
\label{eq:local_channel_update}
\end{equation}
for $k=1,\ldots,K$. The local data blocks are updated as
\begin{equation}
\begin{aligned}
    \mathbf{d}_k^{(i+1)}\!\!\!
    =
    \mathcal{P}_{\mathcal{B}^{D_{\rm d}}}\!
    \Bigg(\!
    \mathbf{d}_k^{(i)}
    +
    \gamma_{d,p}^{(i)}
    \Bigg(
    &
    \frac{\mathbf{E}_k^{\mathsf H}
    \mathbf{H}_{p,k}^{\rm r}
    [\mathbf{h}_{p,k}^{(i)}]^{\mathsf H}
    \mathbf{r}_p^{(i)}}{\sigma_p^2}
    +
    \lambda_d\mathbf{d}_k^{(i)}
    \Bigg)\!
    \Bigg).
\end{aligned}
\label{eq:local_data_update}
\end{equation}
Because all block gradients are evaluated at the same current iterate, the channel and data blocks can be updated in parallel within each satellite. Moreover, the local JCEDD tasks are executed independently across satellites. The step sizes $\gamma_{h,p}^{(i)}$ and $\gamma_{d,p}^{(i)}$ are selected by backtracking to satisfy the adopted sufficient-decrease condition for the local objective in \eqref{eq:local_joint_problem}. In addition, the complex soft-thresholding operator in \eqref{eq:local_channel_update} is applied entrywise as
\begin{equation}
    \mathcal{S}_{\rho}(z)
    =
    \max
    \left\{
    1-\frac{\rho}{|z|},
    0
    \right\}z,
    \label{eq:soft_thresholding}
\end{equation}
with the convention $\mathcal{S}_{\rho}(0)=0$. The projection onto the QPSK box in \eqref{eq:local_data_update} is also applied entrywise:
\begin{equation}
\begin{aligned}
    \mathcal{P}_{\mathcal{B}}(z)
    =
    &
    \min
    \left\{
    \max
    \left\{
    \Re\{z\},
    -\frac{1}{\sqrt{2}}
    \right\},
    \frac{1}{\sqrt{2}}
    \right\}
    \\
    &+
    j
    \min
    \left\{
    \max
    \left\{
    \Im\{z\},
    -\frac{1}{\sqrt{2}}
    \right\},
    \frac{1}{\sqrt{2}}
    \right\}.
\end{aligned}
\label{eq:qpsk_box_projection}
\end{equation}

Although the initial data gradient is zero, the known pilot symbols produce a nonzero initial channel update. After $I_{\rm loc}$ local iterations, define
\begin{align}
    \widehat{\mathbf{h}}_{p,k}^{\rm loc}
    &\triangleq
    \mathbf{h}_{p,k}^{(I_{\rm loc})},\\
    \widehat{\mathbf{d}}_{p,k}^{\rm loc}
    &\triangleq
    \mathbf{d}_k^{(I_{\rm loc})}.
\end{align}
The corresponding stacked estimates are
\begin{align}
    \widehat{\mathbf{h}}_p^{\rm loc}
    &=
    \left[
    (\widehat{\mathbf{h}}_{p,1}^{\rm loc})^{\mathsf T},
    \ldots,
    (\widehat{\mathbf{h}}_{p,K}^{\rm loc})^{\mathsf T}
    \right]^{\mathsf T},\\
    \widehat{\mathbf{d}}_p^{\rm loc}
    &=
    \left[
    (\widehat{\mathbf{d}}_{p,1}^{\rm loc})^{\mathsf T},
    \ldots,
    (\widehat{\mathbf{d}}_{p,K}^{\rm loc})^{\mathsf T}
    \right]^{\mathsf T}.
\end{align}

Each noncentral satellite forwards
\begin{equation}
    \mathcal{U}_p
    \triangleq
    \left\{
    \mathbf{y}_p^{\rm b},
    \widehat{\mathbf{h}}_p^{\rm loc},
    \widehat{\mathbf{d}}_p^{\rm loc}
    \right\}
    \label{eq:uploaded_information}
\end{equation}
to the central satellite, while the central satellite directly retains its own observation and local estimates. Quantization and compression are not considered because the inter-satellite links are assumed to be error-free and capacity-unconstrained.

\subsection{Central Cooperative Refinement}
\label{subsec:central_receiver}

The central satellite jointly refines the satellite-specific channels and the common multiuser data using all satellite observations and local estimates. The channel variables are initialized by the corresponding local estimates:
\begin{equation}
    \mathbf{h}_{p,k}^{(0)}
    =
    \widehat{\mathbf{h}}_{p,k}^{\rm loc}.
    \label{eq:central_channel_initialization}
\end{equation}
The common data vector is initialized by inverse-noise-variance combining:
\begin{equation}
    \mathbf{d}^{(0)}
    =
    \mathcal{P}_{\mathcal{B}^{D}}
    \left(
    \frac{
    \sum_{p=1}^{P}
    \sigma_p^{-2}
    \widehat{\mathbf{d}}_p^{\rm loc}
    }{
    \sum_{p=1}^{P}
    \sigma_p^{-2}
    }
    \right).
    \label{eq:central_data_initialization}
\end{equation}
Thus, local data estimates obtained from satellites with lower noise variance receive larger combining weights.

The central refinement problem is
\begin{equation}
\begin{aligned}
    \min_{\substack{
          \{\mathbf{h}_p\}_{p=1}^{P}\\
          \mathbf{d}\in\mathcal{B}^{D}}}
    &
    \sum_{p=1}^{P}
    \frac{
    \left\|
    \mathbf{r}_p
    [\mathbf{h}_p,\mathbf{d}]
    \right\|_2^2
    }{2\sigma_p^2}
    +
    \mu_h
    \sum_{p=1}^{P}
    \sum_{k=1}^{K}
    \|\mathbf{h}_{p,k}\|_1 -
    \frac{\lambda_d}{2}
    \|\mathbf{d}\|_2^2.
\end{aligned}
\label{eq:central_joint_problem}
\end{equation}
The channel variables remain satellite-specific, whereas the common data vector is inferred jointly from all satellite observations.

At central iteration $i$, the transmitted-signal estimates and residuals are computed using \eqref{eq:iterative_transmitted_signal} and \eqref{eq:iterative_residual}. The channel coefficients are then updated in parallel as
\begin{equation}
\begin{aligned}
    \mathbf{h}_{p,k}^{(i+1)}
    =
    \mathcal{S}_{\gamma_{h,p}^{(i)}\mu_h}
    \left(
    \mathbf{h}_{p,k}^{(i)}
    +
    \frac{\gamma_{h,p}^{(i)}}{\sigma_p^2}
    \mathbf{\Phi}_{p,k}
    [\mathbf{x}_k^{(i)}]^{\mathsf H}
    \mathbf{r}_p^{(i)}
    \right).
\end{aligned}
\label{eq:central_channel_update}
\end{equation}
The common data blocks are updated using the gradients contributed by all satellites:
\begin{equation}
\begin{aligned}
    \mathbf{d}_k^{(i+1)}\!\!\!\!
    =\!
    \mathcal{P}_{\mathcal{B}^{D_{\rm d}}}\!
    \Bigg(\!
    \mathbf{d}_k^{(i)}\!
    +\!
    \gamma_d^{(i)}
    \Bigg(\!
    \sum_{p=1}^{P}
    \frac{\mathbf{E}_k^{\mathsf H}
    \mathbf{H}_{p,k}^{\rm r}
    [\mathbf{h}_{p,k}^{(i)}]^{\mathsf H}
    \mathbf{r}_p^{(i)}}{\sigma_p^2}
    \!+\!
    \lambda_d\mathbf{d}_k^{(i)}
    \!\!\Bigg)\!
    \Bigg).
\end{aligned}
\label{eq:central_data_update}
\end{equation}
The step sizes are selected by backtracking to satisfy the adopted sufficient-decrease condition for the central objective in \eqref{eq:central_joint_problem}. As in the local stage, all forward and adjoint operations are evaluated in a matrix-free manner.

After $I_{\rm cen}$ central iterations, the relaxed data estimates are mapped to the nearest QPSK constellation points:
\begin{equation}
    \widehat{d}_{k,r}
    =
    \arg\min_{c\in\mathcal{C}}
    \left|
    c-d_{k,r}^{(I_{\rm cen})}
    \right|^2,
    \qquad
    r=1,\ldots,D_{\rm d}.
    \label{eq:final_symbol_slicing}
\end{equation}
The final channel estimate for link $(p,k)$ is
\begin{equation}
    \widehat{\mathbf{h}}_{p,k}
    =
    \mathbf{h}_{p,k}^{(I_{\rm cen})}.
    \label{eq:final_channel_estimate}
\end{equation}

\begin{algorithm}[t]
\caption{Low-Complexity Hierarchical JCEDD Receiver}
\label{alg:hierarchical_jcedd}
\begin{algorithmic}[1]
\Require
$\{\mathbf{y}_p^{\rm b},\sigma_p^2\}_{p=1}^{P}$,
$\{\mathbf{p}_k,\mathbf{E}_k\}_{k=1}^{K}$,
candidate dictionaries,
$\mu_h$, $\lambda_d$, $I_{\rm loc}$, $I_{\rm cen}$,
and backtracking parameters
\Ensure
$\{\widehat{\mathbf{h}}_{p,k}\}_{p,k}$ and
$\{\widehat{\mathbf{d}}_k\}_{k=1}^{K}$

\Statex \textbf{Stage 1: Parallel local JCEDD}
\For{$p=1,\ldots,P$ \textbf{in parallel}}
    \State Initialize
    $\mathbf{h}_p^{(0)}=\mathbf{0}$ and
    $\mathbf{d}^{(0)}=\mathbf{0}$
    \For{$i=0,\ldots,I_{\rm loc}-1$}
        \State Compute
        $\{\mathbf{x}_k^{(i)}\}_{k=1}^{K}$ and
        $\mathbf{r}_p^{(i)}$ by \eqref{eq:iterative_transmitted_signal} and
        \eqref{eq:iterative_residual}
        \State Select
        $\gamma_{h,p}^{(i)}$ and $\gamma_{d,p}^{(i)}$
        by backtracking
        \State Update all channels and data using
        \eqref{eq:local_channel_update} and
        \eqref{eq:local_data_update}
    \EndFor
\EndFor
\State Forward $\mathcal{U}_p$ in \eqref{eq:uploaded_information} from each noncentral satellite to the central
satellite; retain the central satellite's own $\mathcal{U}_p$

\Statex \textbf{Stage 2: Central cooperative refinement}
\State Initialize
$\{\mathbf{h}_{p,k}^{(0)}\}_{p,k}$ and $\mathbf{d}^{(0)}$ using
\eqref{eq:central_channel_initialization} and
\eqref{eq:central_data_initialization}

\For{$i=0,\ldots,I_{\rm cen}-1$}
    \State Compute
    $\{\mathbf{x}_k^{(i)}\}_{k=1}^{K}$ and
    $\{\mathbf{r}_p^{(i)}\}_{p=1}^{P}$ by \eqref{eq:iterative_transmitted_signal} and \eqref{eq:iterative_residual}
    \State Select
    $\{\gamma_{h,p}^{(i)}\}_{p=1}^{P}$ and $\gamma_d^{(i)}$
    by backtracking
    \State Update all channel and data blocks using
    \eqref{eq:central_channel_update} and
    \eqref{eq:central_data_update}
\EndFor

\State Obtain
$\{\widehat{d}_{k,r}\}_{k,r}$ and
$\{\widehat{\mathbf{h}}_{p,k}\}_{p,k}$ using
\eqref{eq:final_symbol_slicing} and
\eqref{eq:final_channel_estimate}
\end{algorithmic}
\end{algorithm}

The complete procedure of the proposed hierarchical JCEDD receiver is summarized in Algorithm~\ref{alg:hierarchical_jcedd}.

\subsection{Computational Complexity Analysis}
\label{subsec:complexity}

The proposed receiver reduces the channel search dimension using local beam--delay--Doppler candidate regions and avoids explicitly constructing large sensing and equivalent-channel matrices.

\subsubsection{Arithmetic Complexity}

The beam-domain transformation is performed once per satellite, with aggregate complexity
\begin{equation}
    C_{\rm beam}
    =
    \mathcal{O}
    \left(
    PQN_r(\log N_x+\log N_y)
    \right).
    \label{eq:beam_transform_complexity}
\end{equation}
Let $C_{\rm DD}(Q)$ denote the complexity of applying one DD-domain path operator or its adjoint:
\begin{equation}
    C_{\rm DD}(Q)
    =
    \begin{cases}
        \mathcal{O}(Q), & \text{on-grid delay and Doppler},\\
        \mathcal{O}(Q\log Q), & \text{fractional-grid parameters}.
    \end{cases}
    \label{eq:dd_operator_complexity}
\end{equation}

The per-iteration complexity of local JCEDD at satellite $p$ is given by
\begin{equation}
    C_{{\rm loc},p}^{\rm iter}
    =
    \mathcal{O}
    \left(
    \sum_{k=1}^{K}
    J_{p,k}C_{\rm DD}(Q)
    +
    B_pQ+J_p+D
    \right).
    \label{eq:per_satellite_local_complexity}
\end{equation}
Thus, the aggregate local complexity is
\begin{equation}
    C_{\rm loc}^{\rm iter}
    =
    \sum_{p=1}^{P}C_{{\rm loc},p}^{\rm iter}.
    \label{eq:local_complexity}
\end{equation}

The per-iteration complexity of central refinement is
\begin{equation}
\begin{aligned}
    C_{\rm cen}^{\rm iter}
    =
    \mathcal{O}
    \left(
    \sum_{p=1}^{P}\sum_{k=1}^{K}
    J_{p,k}C_{\rm DD}(Q)
    +
    \sum_{p=1}^{P}(B_pQ+J_p)
    +
    D
    \right).
\end{aligned}
\label{eq:central_complexity}
\end{equation}
The aggregate arithmetic complexity is therefore
\begin{equation}
    C_{\rm total}
    =
    C_{\rm beam}
    +
    I_{\rm loc}C_{\rm loc}^{\rm iter}
    +
    I_{\rm cen}C_{\rm cen}^{\rm iter}.
    \label{eq:total_receiver_complexity}
\end{equation}
Note that the additional objective evaluations caused by backtracking are omitted.

\subsubsection{Comparison with Full-Grid Processing}

Without candidate-region restriction, each link contains $J_{p,k}^{\rm full}=N_rQ$ atoms. A direct atom-wise full-grid matrix-free implementation therefore has per-iteration complexity
\begin{equation}
    C_{\rm full,mf}^{\rm iter}
    =
    \mathcal{O}
    \left(
    PKN_rQ\,C_{\rm DD}(Q)
    \right).
    \label{eq:full_grid_matrix_free_complexity}
\end{equation}
The proposed method replaces $N_rQ$ atoms per link with $J_{p,k}=A_{p,k}T_{p,k}V_{p,k}$ atoms and is thus more efficient when $J_{p,k}\ll N_rQ$.

Explicitly storing $\mathbf{\Phi}_{p,k}[\mathbf{x}]$ and $\mathbf{H}_{p,k}^{\rm r}[\mathbf{h}_{p,k}]$ requires $\mathcal{O}\left(B_pQJ_{p,k}+B_pQ^2  \right)$ complex-valued entries per link, which becomes $\mathcal{O}(N_r^2Q^2)$ under full-grid processing. In contrast, the proposed matrix-free implementation requires $\mathcal{O}(B_pQ+J_p+D)$ working variables at satellite $p$ and $\mathcal{O} \left(    \sum_{p=1}^{P}(B_pQ+J_p)+D\right)$ working variables at the central satellite, excluding candidate parameters and FFT workspace.

\section{Simulation Experiments}
We conduct simulations to evaluate the proposed hierarchical JCEDD  receiver in terms of channel estimation and data detection.

\subsection{Simulation Setup}
\label{subsec:simulation_setup}

\subsubsection{Scenario Configuration}
Unless otherwise specified, we consider a cell-free uplink with $P=3$ cooperating LEO satellites and $K=4$ scheduled single-antenna users. Each satellite employs an $8\times8$ UPA with $N_r=64$ receive antennas. The satellite elevation angle and carrier frequency are set to $50^{\circ}$ and $f_c=2$~GHz, respectively. All results are averaged over $N_{\rm MC}=500$ independent Monte Carlo realizations.

The OTFS grid contains $M=64$ delay bins and $N=64$ Doppler bins, yielding $Q=MN=4096$ DD-domain symbols. With a subcarrier spacing of $\Delta f=15$~kHz, the sampling interval, useful frame duration, and Doppler-bin spacing are $T_s=\frac{1}{M\Delta f}\approx1.0417~\mu\mathrm{s}$, $QT_s\approx4.2667~\mathrm{ms}$, and $\Delta\nu=\frac{1}{QT_s}=234.375~\mathrm{Hz}$, respectively. A reduced CP of four samples is appended, resulting in a total frame duration of approximately $4.2708$~ms.

A $14\times9$ core pilot region containing $126$ symbols is employed. Four delay bins and three Doppler bins are reserved on each side as a protection region. The pilots are independently generated according to $\mathcal{CN}(0,1)$ with unit average power. The remaining $4096-330=3766$ positions carry unit-energy QPSK data symbols.

The signal-to-noise ratio (SNR) at satellite $p$ is defined as the ratio of the noiseless received-signal energy over one OTFS frame to the corresponding expected noise energy, i.e.,
\begin{equation}
    \mathrm{SNR}_p
    \triangleq
    \frac{
    \left\|
    \sum_{k=1}^{K}
    \mathbf{H}_{p,k}\mathbf{x}_k
    \right\|_2^2
    }{
    N_rQ\sigma_p^2
    }.
    \label{eq:satellite_snr_definition}
\end{equation}
The nominal SNR is set to $15$~dB. Satellite-specific offsets of $[0,-1.5,1.0]$~dB are applied, resulting in SNRs of $[15,13.5,16]$~dB for the three satellites.

\subsubsection{Channel Configuration}
The satellite--user channels follow the NTN-CDL-A model~\cite{Shen2022Random}. Each link contains three propagation paths with a root-mean-square (RMS) delay spread of $30$~ns, and the channel remains constant within one OTFS frame. The path-dependent Doppler offset relative to the coarse Doppler center is limited to $200$~Hz. The azimuth and elevation angles of arrival are generated within $[-180^{\circ},180^{\circ}]$ and $[0^{\circ},90^{\circ}]$, respectively.

For each link, the coarse residual-delay center is uniformly located between one and three samples, with a maximum error of $0.4$ sample, while the coarse residual-Doppler center lies within $[-2,2]$ Doppler bins. A $3\times3$ neighborhood around the coarse beam direction is retained, giving $A_{p,k}=9$ candidate beams. The relative delay and Doppler candidate sets are $\{-0.5,0,0.5\}$ samples and $\{-1,-0.5,0,0.5,1\}$ Doppler bins, respectively. Thus, $T_{p,k}=3$, $V_{p,k}=5$, and each link contains $J_{p,k}\!=\!A_{p,k}T_{p,k}V_{p,k}\!=\!135$ candidate beam--delay--Doppler atoms.

\subsubsection{Algorithm Configuration}
The channel sparsity regularization parameter is normalized according to the initial channel-gradient scale. Specifically, for each Monte Carlo realization, we compute $\mu_h=0.15 \max_{p,k} \frac{1}{\sigma_p^2} \left\| \mathbf{\Phi}_{p,k}[\mathbf{p}_k]^{\mathsf H} \mathbf{y}_p^{\rm b} \right\|_{\infty}$. The data regularization parameter is set to $\lambda_d=0.05$. This normalization uses only the received observations, known pilots, and noise variances, without requiring the true channel information. The default numbers of local and central iterations are $I_{\rm loc}=40$ and $I_{\rm cen}=60$, respectively. The initial channel and data step-size scaling factors are both $0.9$. The backtracking contraction factor, maximum number of backtracking trials, and sufficient-decrease tolerance are set to $0.5$, $24$, and $10^{-7}$, respectively.

\begin{figure*}
    \centering
    \subfloat[Channel NMSE versus the number of local iterations.]
    {
        \includegraphics[width=0.47\textwidth]
        {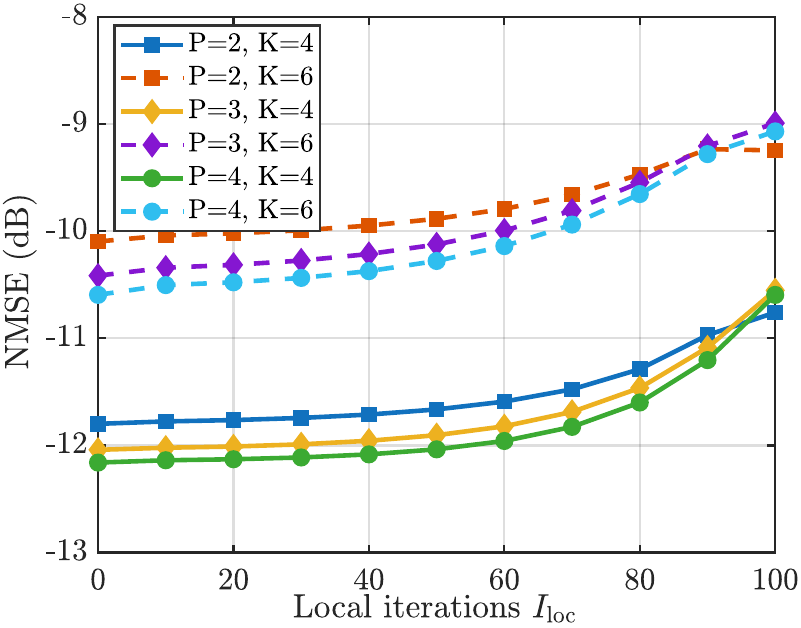}
        \label{fig:iteration_allocation_nmse}
    }
    \hfill
    \subfloat[BER versus the number of local iterations.]
    {
        \includegraphics[width=0.47\textwidth]
        {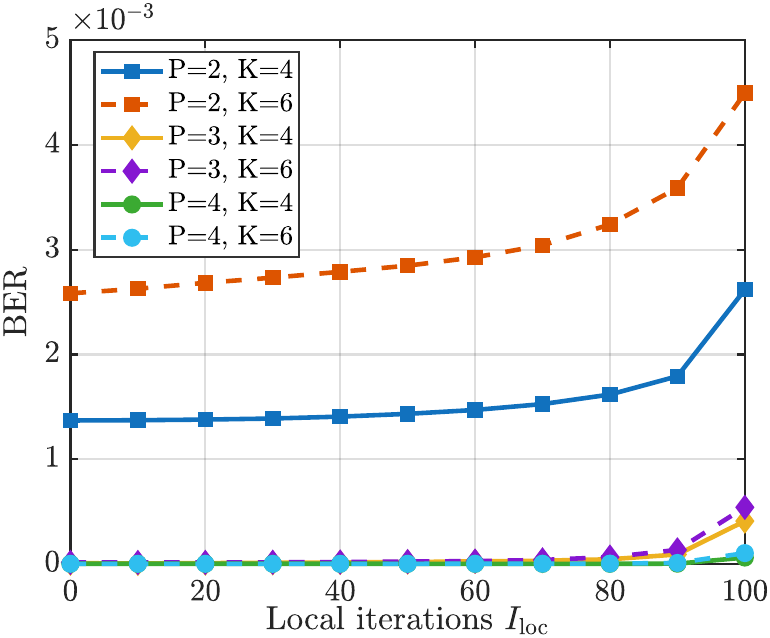}
        \label{fig:iteration_allocation_ber}
    }
    \caption{Channel NMSE and BER versus the number of local iterations for
    different numbers of cooperating satellites and users, where
    $I_{\rm loc}+I_{\rm cen}=100$.}
    \label{fig:iteration_allocation}
\end{figure*}

\subsection{Baselines and Performance Metrics}
\label{subsec:performance_metrics}
The following four baseline receivers are considered.

\begin{itemize}
\item \textbf{Centralized AEP:} The centralized AEP receiver in \cite{Shen2025Massive} uses MRF-BGM-AMP and GAMP to initialize the channel and data estimates, respectively, followed by centralized joint refinement.
\item \textbf{Threshold-based CE + LMMSE:} Each satellite applies the pilot-aided threshold-based channel estimator in \cite{Raviteja2019Embedded}. The central satellite then performs linear minimum mean-square error (LMMSE) data detection using the aggregated observations and channel estimates.
\item \textbf{OMP + LMMSE:} The central satellite jointly estimates the sparse satellite--user channels from all pilot observations using orthogonal matching pursuit (OMP) \cite{Tropp2007Signal}, followed by LMMSE data detection.
\item \textbf{MB-ULMO + LMMSE:} Each satellite applies MB-ULMO \cite{Jafri2026Bayesian} with multiple-measurement-vector Bayesian learning (MMV-BL) for local multiuser channel estimation, followed by centralized LMMSE data detection.
\end{itemize}

Channel-estimation and data-detection performance are evaluated using the normalized mean-square error (NMSE) and bit error rate (BER), respectively. For the $m$th realization, let $\mathbf{H}_{p,k}^{\rm r,(m)} \triangleq [(\mathbf{S}_p^{(m)}\mathbf{U}_{\rm a}^{\mathsf H}) \otimes\mathbf{I}_Q]\mathbf{H}_{p,k}^{(m)}$ denote the true equivalent channel of link $(p,k)$ over the retained beam set, and let $\widehat{\mathbf{H}}_{p,k}^{\rm r,(m)}\triangleq\mathbf{H}_{p,k}^{\rm r}[\widehat{\mathbf{h}}_{p,k}^{(m)}]$ denote its reconstruction from the estimated coefficients. The reported channel NMSE in dB is
\begin{equation}
    \textstyle \mathrm{NMSE}
    \!=\!
    10\log_{10}\!
    \left(\!
    \frac{1}{N_{\rm MC}}
    \sum_{m=1}^{N_{\rm MC}}
    \frac{
    \sum_{p=1}^{P}\sum_{k=1}^{K}
    \left\|
    \widehat{\mathbf{H}}_{p,k}^{\rm r,(m)}
    -
    \mathbf{H}_{p,k}^{\rm r,(m)}
    \right\|_F^2
    }{\sum_{p=1}^{P}\sum_{k=1}^{K}
    \left\|
    \mathbf{H}_{p,k}^{\rm r,(m)}
    \right\|_F^2
    }\!
    \right)\!\!.
    \label{eq:average_channel_nmse}
\end{equation}

For BER evaluation, let $b_{k,r,q}^{(m)}\in\{0,1\}$ and $\widehat{b}_{k,r,q}^{(m)}\in\{0,1\}$ denote the $q$th transmitted and detected bits, respectively, associated with the $r$th QPSK symbol of user $k$, where $q\in\{1,2\}$. The reported BER is
\begin{equation}
    \!\!\!\mathrm{BER}
    \!=\!
    \frac{1}{2KD_{\rm d}N_{\rm MC}}
    \sum_{m=1}^{N_{\rm MC}}
    \sum_{k=1}^{K}
    \sum_{r=1}^{D_{\rm d}}
    \sum_{q=1}^{2}
    \mathbb{I}
    \left\{
    \widehat{b}_{k,r,q}^{(m)}
    \neq
    b_{k,r,q}^{(m)}
    \right\}\!.
    \label{eq:average_ber}
\end{equation}

\subsection{Impact of Iteration Allocation}

We consider $P\in\{2,3,4\}$ cooperating satellites and $K\in\{4,6\}$ scheduled users. The total iteration budget is fixed at $I_{\rm loc}+I_{\rm cen}=100$, while $I_{\rm loc}$ varies from $0$ to $100$, thereby characterizing the tradeoff between local processing and central refinement.

As shown in Fig.~\ref{fig:iteration_allocation}, allocating a small or moderate number of iterations to the local stage preserves most of the performance and provides an informative initialization for central refinement. However, both NMSE and BER deteriorate when excessive iterations are assigned to local processing, particularly for $I_{\rm loc}\gtrsim70$--$80$. This indicates that independently processing single-satellite observations cannot fully exploit the consistency of the common data symbols across satellites. Therefore, sufficient central iterations should be retained for cooperative refinement. When $I_{\rm loc}=100$, the central stage is removed, leading to a clear performance degradation.

Increasing $P$ substantially improves data detection, especially in the more heavily loaded case with $K=6$. For $P=4$, the BER remains close to zero over most iteration allocations. The improvement in channel NMSE is less pronounced because the channels are satellite-specific, whereas the transmitted data are common to all satellite observations and directly benefit from cooperative combining. Increasing $K$ degrades both metrics because of stronger multiuser interference and the larger number of jointly estimated variables.

\begin{figure*}
    \centering
    \subfloat[BER versus SNR.]
    {
        \includegraphics[width=0.47\textwidth]
        {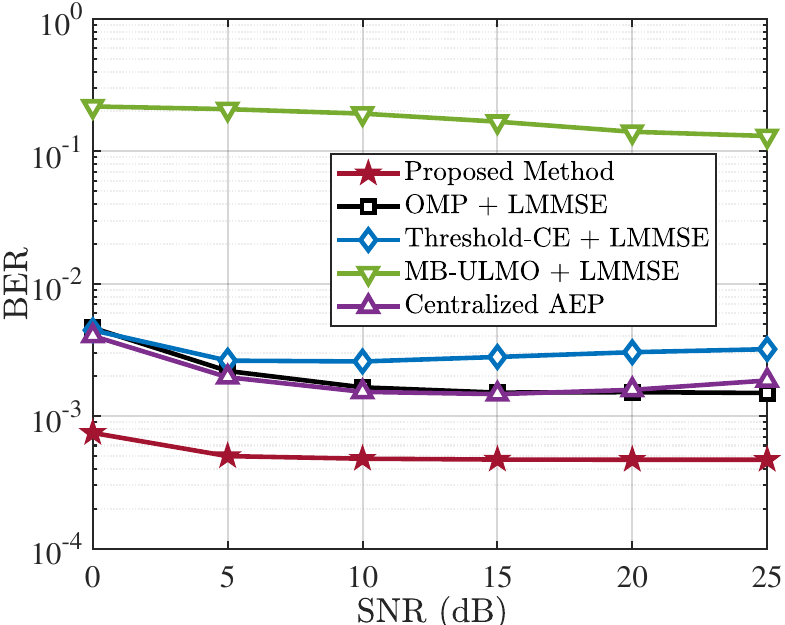}
        \label{fig:ber_vs_snr}
    }
    \hfill
    \subfloat[Channel NMSE versus SNR.]
    {
        \includegraphics[width=0.47\textwidth]
        {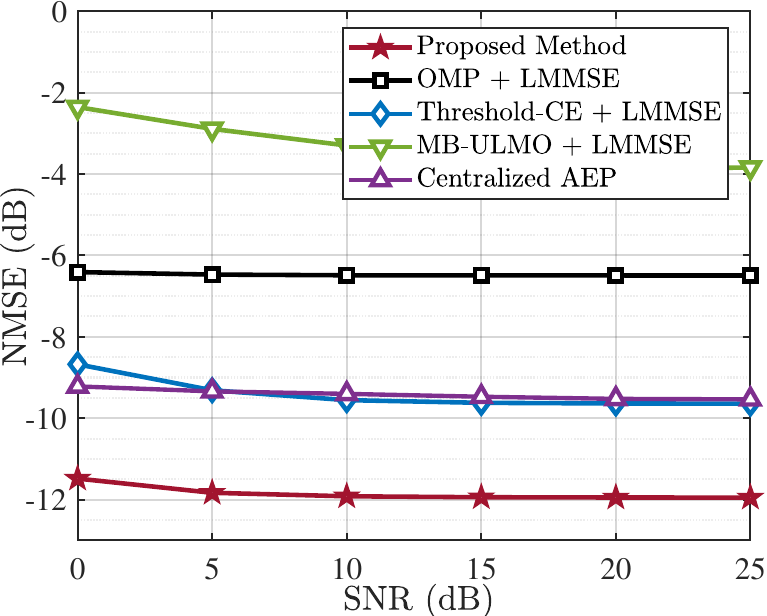}
        \label{fig:nmse_vs_snr}
    }

    \subfloat[BER versus the number of cooperating satellites.]
    {
        \includegraphics[width=0.47\textwidth]
        {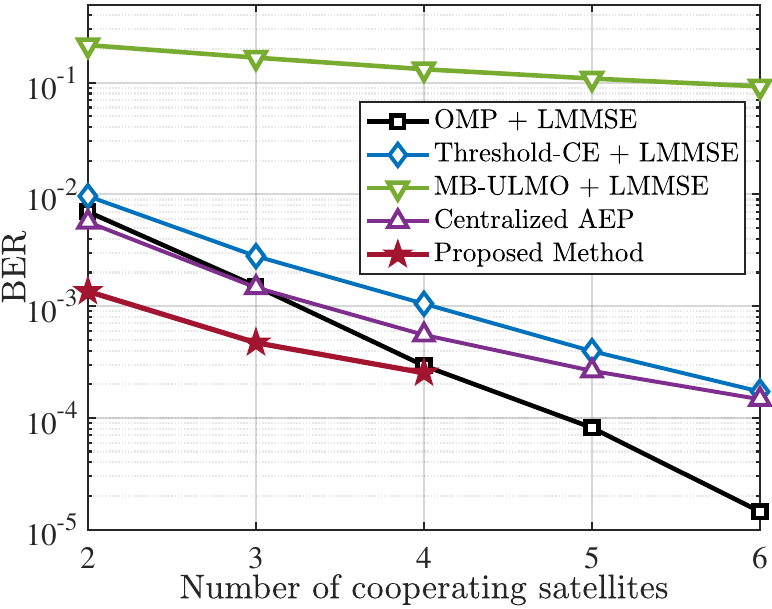}
        \label{fig:ber_vs_satellite_count}
    }
    \hfill
    \subfloat[Channel NMSE versus the number of cooperating satellites.]
    {
        \includegraphics[width=0.47\textwidth]
        {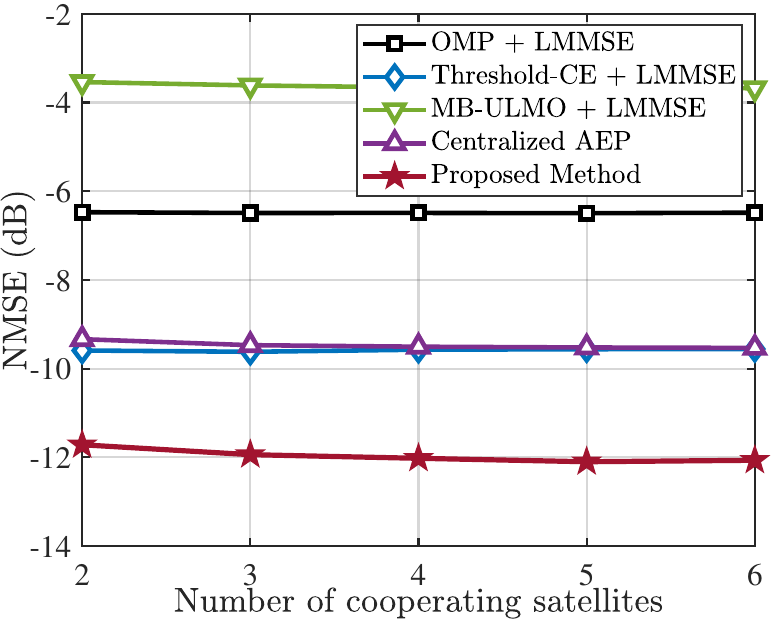}
        \label{fig:nmse_vs_satellite_count}
    }
    \caption{BER and channel NMSE under different SNRs and numbers of
    cooperating satellites.}
    \label{fig:performance_comparison}
\end{figure*}

\subsection{Performance Comparison Under Different Scenarios}

Fig.~\ref{fig:performance_comparison} compares the proposed receiver with the baselines under different SNRs and numbers of cooperating satellites. In both scenarios, the proposed receiver consistently achieves the lowest BER and channel NMSE, demonstrating that its performance advantage is maintained under different noise levels and cooperation scales.

As shown in Figs.~\ref{fig:ber_vs_snr} and \ref{fig:nmse_vs_snr}, increasing the SNR reduces its BER from approximately $7\times10^{-4}$ to $4\times10^{-4}$ and improves its NMSE from about $-11.6$~dB to $-12$~dB. Compared with the best baseline, the proposed receiver provides an NMSE gain of approximately $2.4$~dB together with a clear BER reduction. The gradual performance saturation at moderate-to-high SNRs is mainly caused by beam-domain candidate-region truncation. Since the angles of arrival are generally off-grid, the channel energy spreads across multiple beams. As each estimated link is restricted to its own candidate region, the true channel components outside this region cannot be represented, resulting in a residual structural error even at high SNR.

To ensure a fair comparison across different numbers of cooperating satellites, the SNR at every satellite is fixed at $15$~dB. As shown in Fig.~\ref{fig:ber_vs_satellite_count}, increasing $P$ substantially improves data detection. The BER of the proposed receiver decreases from approximately $1.4\times10^{-3}$ for $P=2$ to about $2.5\times10^{-4}$ for $P=4$, while no bit errors are observed for $P\geq5$ under the adopted Monte Carlo budget. In contrast, Fig.~\ref{fig:nmse_vs_satellite_count} shows only a modest NMSE improvement, from approximately $-11.7$~dB to $-12.1$~dB. This is because each additional satellite provides a new observation but also introduces a new set of satellite-specific channels to be estimated. Nevertheless, the more accurate common data estimate enabled by multi-satellite cooperation reduces error propagation during central joint refinement, thereby slightly improving channel estimation.

\section{Conclusion}
This paper investigated JCEDD for multiuser multi-LEO-satellite cell-free OTFS uplinks. We established a signal model based on local single-satellite observations and formulated JCEDD as a structured bilinear inference problem involving satellite--user-specific sparse beam--delay--Doppler channels and a common multiuser data vector. A hierarchical receiver was developed, in which each satellite performs local JCEDD and the central satellite jointly refines the link-specific channels and common data using the aggregated observations and local estimates. Coarse-information-aided candidate regions and matrix-free forward and adjoint operations were employed to reduce the channel search dimension and avoid constructing large sensing and equivalent-channel matrices. Simulation results showed that the proposed receiver outperforms representative baselines in both channel estimation and data detection. The proposed framework provides a promising low-complexity solution for reliable uplink reception in future 6G non-terrestrial networks. Future work will investigate the impact of capacity-limited inter-satellite links on the communication overhead and performance of hierarchical JCEDD.

\newpage
\ifCLASSOPTIONcaptionsoff
  \newpage
\fi

\balance
\bibliographystyle{IEEEtran}
\bibliography{./bib/Refabrv,./bib/IEEEbib1}

\end{document}